\documentclass[11pt]{article} 

\usepackage[utf8]{inputenc} 

\usepackage{geometry} 
\usepackage{graphicx} 

\usepackage{graphicx}
\usepackage{caption}
\usepackage{epstopdf}
\usepackage{amsmath} 
\usepackage{amsfonts}
\usepackage{booktabs} 
\usepackage{array} 
\usepackage{paralist} 
\usepackage{verbatim} 
\usepackage{subfig}
\usepackage{bm}
\usepackage{fancyhdr} 
\usepackage{sectsty}
\allsectionsfont{\sffamily\mdseries\upshape} 

\usepackage[nottoc,notlof,notlot]{tocbibind} 
\usepackage[titles,subfigure]{tocloft} 

\title{Chaos in Homeostatically Regulated Neural Systems} 

\author{Wilten Nicola, Peter Hellyer, Sue Ann Campbell, Claudia Clopath} 
\date{\today} 

\begin{document}
\maketitle    

\section*{Abstract}
Low-dimensional yet rich dynamics often emerge in the brain.  Examples include oscillations and chaotic dynamics during sleep, epilepsy, and voluntary movement.   However, a general mechanism for the emergence of low dimensional dynamics remains elusive.  Here, we consider Wilson-Cowan networks and demonstrate through numerical and analytical work that a type of homeostatic regulation of the network firing rates can paradoxically lead to a rich dynamical repertoire.  The dynamics  include mixed-mode oscillations, mixed-mode chaos, and chaotic synchronization.  This is true for single recurrently coupled node, pairs of reciprocally coupled nodes without self-coupling, and networks coupled through experimentally determined weights derived from functional magnetic resonance imaging data.  In all cases, the stability of the homeostatic set point is analytically determined or approximated.  The dynamics at the network level are directly determined by the behavior of a single node system through synchronization in both oscillatory and non-oscillatory states.  Our results demonstrate that rich dynamics can be preserved under homeostatic regulation or even be caused by homeostatic regulation.

\section{Introduction} 

The human brain contains billions of neurons each receiving potentially thousands of connections from their neighbours.   Despite this complexity, low-dimensional dynamics often appear in the brain in different regions and contexts.   Examples include oscillations such as the theta and gamma oscillations in the hippocampus \cite{theta,gamma,theta2}, low dimensional oscillatory dynamics during grasping and other motions \cite{churchland}, or even low dimensional chaotic dynamics during epileptic seizures and different sleep phases \cite{destexhe}.  These dynamics are sometimes pathological, such as during epileptic seizures while other times they are functional, such as during sleep states.  Despite the low-dimensionality, the dynamics these systems display are often complex \cite{destexhe}.  However, a general mechanism as to how these dynamical regimes might initially emerge remains elusive.  

If these dynamical regimes are indeed learned and not inherited, plasticity in the synaptic weights that couple neurons together is necessary.  For many neural circuits, strong evidence exists for a form of homeostatic plasticity  \cite{froemke,homeostasis2,homeostasis3,homeostasis4}.  The function of homeostatic plasticity is to prevent run-away excitation in the circuit and thus pathological states such as epileptic seizures.  Additionally, homeostatic plasticity prevents a catastrophic loss of neuronal activity which results in network quiescence.  In other words, homeostatic plasticity serves to maintain a stable background firing rate.  

Recent modeling work has demonstrated a novel inhibitory homeostatic plasticity mechanism designed to regulate activity \cite{vogels}.  This mechanism works by applying slow variations in the synaptic weights from the inhibitory neurons to the excitatory neurons \cite{vogels}.  As the excitatory neurons start firing in excess of their homeostatic set points, the synaptic weights from the inhibitory neurons increase in strength to prevent run-away excitation.  If the excitation in the network is too low, the inhibitory weights decrease in strength to disinhibit the excitatory neurons.   The homeostatic mechanism can drive initially synchronized activity into the asynchronous irregular regime defined by variable spiking but with a constant time averaged firing rate \cite{vogels,brunel}.  

These homeostatic mechanisms fundamentally exist to stabilize network dynamics to an equilibrium point \cite{homeostasis4}.  Indeed, they exist as a counter mechanism to offset the often destabilizing effects of Hebbian plasticity \cite{homeostasis4}.  Thus, it is surprising to consider homeostasis to be the potential source of complex dynamical systems.  However, recent work has suggested that these homeostatic mechanisms yield rich dynamics in large networks \cite{Peter}.  For example a, coupled Wilson-Cowan (or mean-field) system with inhibitory homeostatic synaptic plasticity and excitatory weights estimated from functional Magnetic Resonance Imaging (fMRI) data showed rich spontaneous dynamics such as neuronal avalanches \cite{Peter}.  However, it is difficult to determine what the source of the rich dynamical repertoire of these systems is as the underlying networks contain neuronal noise, synaptic transmission delays, non-smooth dynamics, and complex coupling.  All four components may be the source of complex dynamics.  

In this work, we attempt to disentangle what effect the homeostatic dynamics have by analyzing a smooth Wilson-Cowan (\cite{WilsonCowan}) system similar to the system numerically analyzed in \cite{Peter} without delays or noise.  We show that the rich dynamical repertoire of these macroscopic networks is an intrinsic component of homeostasis and thus not dependent on any other network conditions such as noise or transmission delays.  Indeed, complex dynamics arise in a single node with recurrent excitation and homeostatically regulated inhibition.  For example, the single node system displays a period doubling cascade to chaos, mixed-mode oscillations, and mixed-mode chaos.  Furthermore, we demonstrate that these results also occur in coupled dual node systems, and in large coupled node systems.  The coupling in the large network is identical to the connectivity considered in \cite{Peter} and derived from functional magnetic resonance imaging data from \cite{hagmann,Honey}.  For both cases, we find that the complex dynamics of the single node carry over to higher dimensions.  Finally, we consider node and connection deletion in simulations using the data derived coupling matrices.  We find that the homeostatic effect on firing rate stability is substantially boosted by the deletion of very specific nodes or connections in the network.

\section{Materials and Methods} 
The system of equations we consider phenomenologically model the average activity of a population of neurons \cite{WilsonCowan}.  The population consists of a subpopulations of excitatory neurons, $E$, and inhibitory neurons, $I$.  Each population corresponds to a single equation governed by the following dynamical system:
\begin{eqnarray}
\tau_E E' &=& -E + \phi\left(W^{EE} E - W^{EI}I\right) \label{WC1}\\
\tau_I I' &=& -I + \phi\left(W^{IE}E \right)\label{WC2}
\end{eqnarray}
The coupling terms $W^{EE}, W^{EI}, W^{IE}$ are all assumed to be positive while the self-inhibition term is assumed to be zero, for simplicity.  The function $\phi(x)$ is a sigmoidal transfer function that transforms the net current arriving at a population into the population activity.  The time constants $\tau_E$ and $\tau_I$ denote time scales of the excitatory and inhibitory populations, respectively.  The equations (\ref{WC1})-(\ref{WC2}) are more commonly referred to as the Wilson-Cowan system \cite{WilsonCowan}.  Here, we also consider the homeostatic modification from \cite{vogels,Peter}: 
\begin{eqnarray}
\tau_W {W^{EI}}' = I(E-p) \label{WC3}
\end{eqnarray}
where $p$ is the homeostatic set point for the networks excitatory activity.  Equation (\ref{WC3}) alters the dynamics of the $EI$ inhibitory synaptic weight in order to drive the excitatory population toward $p$, the homeostatic set point of the network.   Equations (\ref{WC1})-(\ref{WC3}) together define the dynamics of a single, recurrently coupled node.  As we will see in Section  \ref{secSN}, analyzing the single node system is vital towards understanding the dynamics of the large network.  

The network equations are given by the following: 
\begin{eqnarray}
\tau_E E_k' &=& -E_k + \phi\left(\sum_{i=1}^N W_{ik}^{EE} E_i - W^{EI}_k I_k\right) \label{fn1}\\ 
\tau_I I_k' &=& - {I_k} + \phi(W_{k}^{IE} E_k)\label{fn2} \\
\tau_W{W^{EI}_k} ' &=&  I_k (E_k - p)  \label{fn3}
\end{eqnarray}
The excitatory activity of population $k$ is given by $E_k$ while the inhibitory activity is given by $I_k$ for $k=1,2,\ldots N$.   These nodes are coupled by the potentially long range weight projection matrix $\bm W^{EE}$ while a node inhibits itself through the diagonal weight matrix $\bm{W}^{EI}$.  We assume that no long-range inhibition is possible, hence the diagonal nature of $\bm{W}^{EI}$.   The time constants for the excitatory, inhibitory, and inhibitory homeostatic synaptic weight are given by $\tau_E$, $\tau_I$, and $\tau_W$, respectively.    Furthermore, we will assume that a node can only excite its own inhibitory population. and thus $\bm W^{IE}$ is also diagonal.  

The transfer function $\phi(x)$ is a smooth sigmoid function which we will constrain to satisfy the following properties: 
\begin{eqnarray}
\phi'(x) &>& 0, \forall x \label{prop1}\\
\lim_{x\rightarrow \infty} \phi(x) &=& 1 \\
\lim_{x\rightarrow -\infty} \phi(x) &=& 0 \label{prop3}
\end{eqnarray}
While our derivations and analysis are general for sigmoid functions that satisfy (\ref{prop1})-(\ref{prop3}), we consider the logistic function: 
\begin{eqnarray}
\phi(x) = \frac{1}{1+\exp(-ax)}, \quad \phi'(x) = a\phi(x)(1-\phi(x))
\end{eqnarray}
for numerical applications.  The parameter $a$ determines the steepness of the sigmoid.   While $\phi(x)$ is a smooth sigmoid function, other transfer functions are also possible.  In particular, various non-smooth variants of $\phi(x)$ can also be considered with differing effects on the final dynamics of the network \cite{Bard1,Wilten1}.  We leave this for future work.  

To simplify the notation further, we will rescale time with $\hat{t} = \tau_I t$.  For the single node, this yields the following system: 
\begin{eqnarray}
\tau_1 E' &=&-E + \phi(W^E E - W^II) \label{sn1} \\\
 I' &=& - I + \phi(\theta E) \\
\tau_2 {W^I}' &=& I (E-p) \label{sn3}.
\end{eqnarray}
with $\tau_1=\tau_E/\tau_I$, $\tau_2 = \tau_W/\tau_I$.  For simplicity, we will relabel the scalar parameters in the single and dual node cases with $W^E$ and $W^I$ for $EE$ and $EI$ synaptic weights and $\theta$ for the $IE$ synaptic weight.  Finally, the coupling matrix for the large network, $\bm{W}^{EE}$, is  derived from functional neural imaging data (see \cite{Peter,hagmann,Honey} for further details).    These data-derived coupling matrices have no self-coupling between nodes ($\bm W^{EE}_{ii} = 0$).  This would seem to imply that analysis of the single system driven by self coupling given by equations (\ref{WC1})-(\ref{WC3}) does not help in understanding the dynamics of the full network where $W^{EE}_{ii}=0, \forall i$.  However, as we will see the symmetric double-node system without self-coupling has largely identical dynamics to the single-node system: 
\begin{eqnarray}
\tau_1 E_1' &=&-E_1 + \phi(W^E E_2 - W^I_1 I_1) \label{dn1} \\\
 I_1' &=& - I_1 + \phi(\theta E_1) \\
\tau_2 {W^I_1}' &=& I_1 (E_1-p) \\
\tau_1 E_2' &=&-E_2 + \phi(W^E E_1 - W^I_2I_2)  \\\
 I_2' &=& - I_2 + \phi(\theta E_2) \\
\tau_2 {W^I_2}' &=& I_2 (E_2-p) \label{dn6}
\end{eqnarray}
and in fact synchronizes to solutions of the single-node system.     

The parameter values we consider for all systems are shown in Table \ref{TAB1}, unless otherwise specified as a bifurcation parameter or the Figures.

We structure the paper as follows:  In Section \ref{secSN} we analyze the single-node system and demonstrate that the majority of the rich dynamics we see for both the dual node and the full network are present for the single node.  In Section \ref{secDN} we numerically demonstrate that the dual node system without self-coupling synchronizes to the single node system analyzed in Section \ref{secDN}.  Finally, in Section \ref{secFN}, we simulate and analyze the full network equations demonstrating a direct inheritance of their dynamics from the single node system.  

\section{Single Node Analysis} \label{secSN}
 
\subsection{Local Analysis} 
Due to the homeostatic mechanism in equation (\ref{WC3}), only one equilibrium exists and is determined by the following equations: 
\begin{eqnarray} \overline{E} = p, \quad \overline{I} = \phi( \theta p), \quad  \overline{W^I} = \frac{W^Ep - \phi^{-1}(p)}{\phi(\theta p )}
\end{eqnarray}  
which is valid for $p\in(0,1)$.   We will subsequently refer to this equilibrium as $\bar{\bm x}=(\overline{E},\overline{I},\overline{W^I})$.   As $W^I>0$ we require:
$$ W^E p > \phi^{-1}(p).$$ 
This sets a range on the admissable values of $W^E$ allowed as a function of $p$, in addition to the constraint that $W^E>0$. Note that these two inequalities coincide when $\phi^{-1}(p) = 0$.  For our sigmoid, this implies that we can consider $p<0.5$ and thus all $W^E>0$. 

After some simplification, the Jacobian of this system is given by 
\begin{eqnarray}
J = \begin{pmatrix} -\frac{1}{\tau_1} + \frac{\phi'(\phi^{-1}(p))W^E}{\tau_1} & -\frac{\overline{W^I} \phi'(\phi^{-1}(p))}{\tau_1} & -\frac{\overline{I}\phi'(\phi^{-1}(p))}{\tau_1}  \\  \phi'(\theta p )\theta & -1& 0 \\ \frac{\overline{I}}{\tau_2} & 0 & 0       \end{pmatrix} .
\end{eqnarray} 
Which yields the following characteristic polynomial for the single node system:
\begin{eqnarray}
C_{SN}(\lambda) &=& \lambda^3 + \lambda^2 \left(\frac{1-W^E\phi'(\phi^{-1}(p))}{\tau_1} + 1\right) + \lambda\left(\frac{1-W^E\phi'(\phi^{-1}(p))}{\tau_1} +\frac{ \overline{W^I} \phi'(\phi^{-1}(p))\phi'(\theta p)\theta}{\tau_1}+\frac{\overline{I}^2 \phi'(\phi^{-1}(p))}{\tau_1\tau_2}\right)\nonumber\\&+&\frac{\overline{I}^2 \phi'(\phi^{-1}(p))}{\tau_1\tau_2}.
\end{eqnarray}
The determinant of the Jacobian is given by 
\begin{eqnarray}
\det{J} = \lambda_1\lambda_2\lambda_3 = -\frac{\overline{I}^2\phi'(\phi^{-1}(p))}{\tau_1\tau_2}= -\frac{\phi(\theta p)^2\phi'(\phi^{-1}(p))}{\tau_1\tau_2}<0.
\end{eqnarray}
This would seem to immediately limit the dynamical repertoire of this system and is at the root of core functionality of the homeostatic variable.   Indeed, due to the dynamics of ${W^I}'$, aside from $(\overline{E},\overline{I},\overline{W^I})$, no other equilibria exist and thus local bifurcations that create or destroy equilibria via $\lambda =0$ crossings are not possible.  This implies that no bistability in equilibria is possible, as in other classical Wilson-Cowan systems. Thus, we can attempt to look for Hopf bifurcations.  Furthermore, as the system is cubic and the determinant is negative, one of the eigenvalues is always negative.  This corresponds to the existence of a stable manifold for the equilibrium globally in the parameter space.  The other eigenvalues must both be real and of the same sign, or complex.   

To determine the potential loss of stability due to Hopf-bifurcations, substitution of $\lambda = i\omega$ into the characteristic polynomial yields the following:
\begin{eqnarray*}
0&=&-i\omega^3-\omega^2 \left(\frac{1-W^E\phi'(\phi^{-1}(p))}{\tau_1} + 1\right) +i\omega \left(\frac{1-W^E\phi'(\phi^{-1}(p))}{\tau_1} +\frac{ \overline{W^I} \phi'(\phi^{-1}(p))\phi'(\theta p)\theta}{\tau_1}+\frac{\overline{I}^2 \phi'(\phi^{-1}(p))}{\tau_1\tau_2}\right) \\&+&\frac{\overline{I}^2 \phi'(\phi^{-1}(p))}{\tau_1\tau_2},
\end{eqnarray*}
which after equating real and imaginary parts yields
\begin{eqnarray}
0&=&\omega^3 - \omega  \left(\frac{1-W^E\phi'(\phi^{-1}(p))}{\tau_1} +\frac{ \overline{W^I} \phi'(\phi^{-1}(p))\phi'(\theta p)\theta}{\tau_1}+\frac{\overline{I}^2 \phi'(\phi^{-1}(p))}{\tau_1\tau_2}\right) \\
0 &=& \omega^2 \left(\frac{1-W^E\phi'(\phi^{-1}(p))}{\tau_1} + 1\right)- \frac{\overline{I}^2 \phi'(\phi^{-1}(p))}{\tau_1\tau_2}.
\end{eqnarray}
Solving for $\omega$ as a function of the network parameters yields:
\begin{eqnarray}
\omega &=&\sqrt{\frac{1-W^E\phi'(\phi^{-1}(p))}{\tau_1} +\frac{ \overline{W^I} \phi'(\phi^{-1}(p))\phi'(\theta p)\theta}{\tau_1}+\frac{\overline{I}^2 \phi'(\phi^{-1}(p))}{\tau_1\tau_2}}.
\end{eqnarray}
The Hopf bifurcation curve is implicitly defined by 
\begin{eqnarray*}
0=\left(\frac{1-W^E\phi'(\phi^{-1}(p))}{\tau_1} +\frac{ \overline{W^I} \phi'(\phi^{-1}(p))\phi'(\theta p)\theta}{\tau_1}+\frac{\overline{I}^2 \phi'(\phi^{-1}(p))}{\tau_1\tau_2}\right)\left(\frac{1-W^E\phi'(\phi^{-1}(p))}{\tau_1} + 1\right)-\frac{\overline{I}^2 \phi'(\phi^{-1}(p))}{\tau_1\tau_2}
\end{eqnarray*}
Defining the following quantities
\begin{eqnarray}
\mu &=& \frac{1-W^E\phi'(\phi^{-1}(p))}{\tau_1}\\
F(\theta) &=& \frac{1-p^{-1}\phi^{-1}(p)\phi'(\phi^{-1}(p))}{\tau_1}\\
\kappa(\theta) &=& \frac{p\phi'(\theta p) \theta }{\phi(\theta p) }\\
D(\theta) &=& \frac{\overline{I}^2 \phi'(\phi^{-1}(p))}{\tau_1\tau_2},
\end{eqnarray}
then the Hopf bifurcation condition can be written as a quadratic equation in $\mu$.  Solving for $\mu$ yields
\begin{eqnarray}
\mu_\pm = \frac{-(D(\theta)+F(\theta)\kappa(\theta)+1-\kappa(\theta)) \pm \sqrt{ (F(\theta)\kappa(\theta)+D(\theta)+1-\kappa(\theta))^2 - 4 \kappa(\theta)F(\theta)(1-\kappa(\theta)}}{2(1-\kappa(\theta))}  .
\end{eqnarray}
Only the positive branch of $\mu$ yields a definite Hopf-bifurcation as we require $\omega^2 =\mu_\pm(1-\kappa(\theta))+F(\theta)\kappa(\theta) +D(\theta)>0$ 
\begin{eqnarray}
\omega_{\pm}^2 = \frac{-(1-\kappa(\theta) - \kappa(\theta)F(\theta) - D(\theta))^2 \pm \sqrt{  \left(1-\kappa(\theta) - \kappa(\theta)F(\theta) - D(\theta)\right)^2+4(1-\kappa(\theta))D(\theta)}}{2}>0
\end{eqnarray}
which implies that $\mu_-<0$ is thus an inadmissable solution for a Hopf-bifurcation while $\mu_+$ is an admissable under the sufficient condition
\begin{eqnarray} \kappa(\theta)=\frac{p\phi'(\theta p) \theta}{\phi(\theta p)}<1\label{gc1}\end{eqnarray}  
By considering the properties of the sigmoid function $\phi(x)$, a routine derivation shows that the inequality (\ref{gc1}) holds when $a<\left(p^2(1-\phi(\theta p)\right)^{-1}$ or more colloquially, when the sigmoid is not too sharp.   The final Hopf bifurcation curve is given by:
\begin{eqnarray}
W^E_{Hopf}(\theta) = \frac{1}{\phi'(\phi^{-1}(p))}\left(1-\tau_1 \mu_+(\theta)\right) \label{Hopf}.
\end{eqnarray}
in the $(\theta,W^E)$ parameter space.

Given the fact that we can explicitly solve for the Hopf-bifurcation curve, we can simulate in its vicinity to determine the resulting behavior of the single-node system.   Direct numerical simulation in addition to numerical continuation using XPPAUT (not shown) indicate that the Hopf bifurcation is likely supercritical, as stable limit cycles emerge for $W^E>W^E_{Hopf}(\theta)$ (Figure \ref{FIG1}, \ref{FIG1}A).   Computing the first Lyapunov coefficient is cumbersome for the full-3D system as it requires a center manifold reduction.  However, for $\theta = 0$ case, one can prove that the Lyapunov coefficient is strictly negative (see Supplementary Section S1).  Thus, we should expect that the first Lyapunov exponent is negative for small $\theta$ which suggests a supercritical Hopf bifurcation (see Supplementary Material S2).

Finally, taking the limits $\theta \rightarrow 0$ or $\theta \rightarrow \infty$ yields
\begin{eqnarray}
W^{E}_{Hopf}(0) =W^{E}_{Hopf}(\infty) =  \frac{1}{\phi'(\phi^{-1}(p))} \label{eqlimit}
\end{eqnarray}
with $W^E_{Hopf}(\theta)\geq W^E_{Hopf}(0)$.  The inequality can be proven by considering that $F(\theta)\geq 0$, $\mu_+(\theta)\leq 0$ where equality only occurs in the asymptotic limits considered in (\ref{eqlimit}). The value $W^E_{Hopf}(0)$ is the critical value after which synaptic homeostasis can no longer guarantee stability of the equilibrium $\bar{\bm x}$.  After this value, depending on the strength of the excitatory to inhibitory coupling $\theta$, stability is lost through a supercritical Hopf bifurcation.  This is however not a catastrophic bifurcation, and thus near the onset of the Hopf bifurcation we are still confined to a neighbourhood around $\bar{\bm x}$.  Note that for the sigmoid we consider, $W^E_{Hopf} = \frac{1}{ap(1-p)}$,  which implies that smoother sigmoids (small $a$) yield a larger parameter region of homeostatic control.  

\subsection{Period Doubling Cascade to Chaos Followed by a Pinching of the Tent Map} 

\begin{figure}[htp!]
\centering
\includegraphics[scale=0.9]{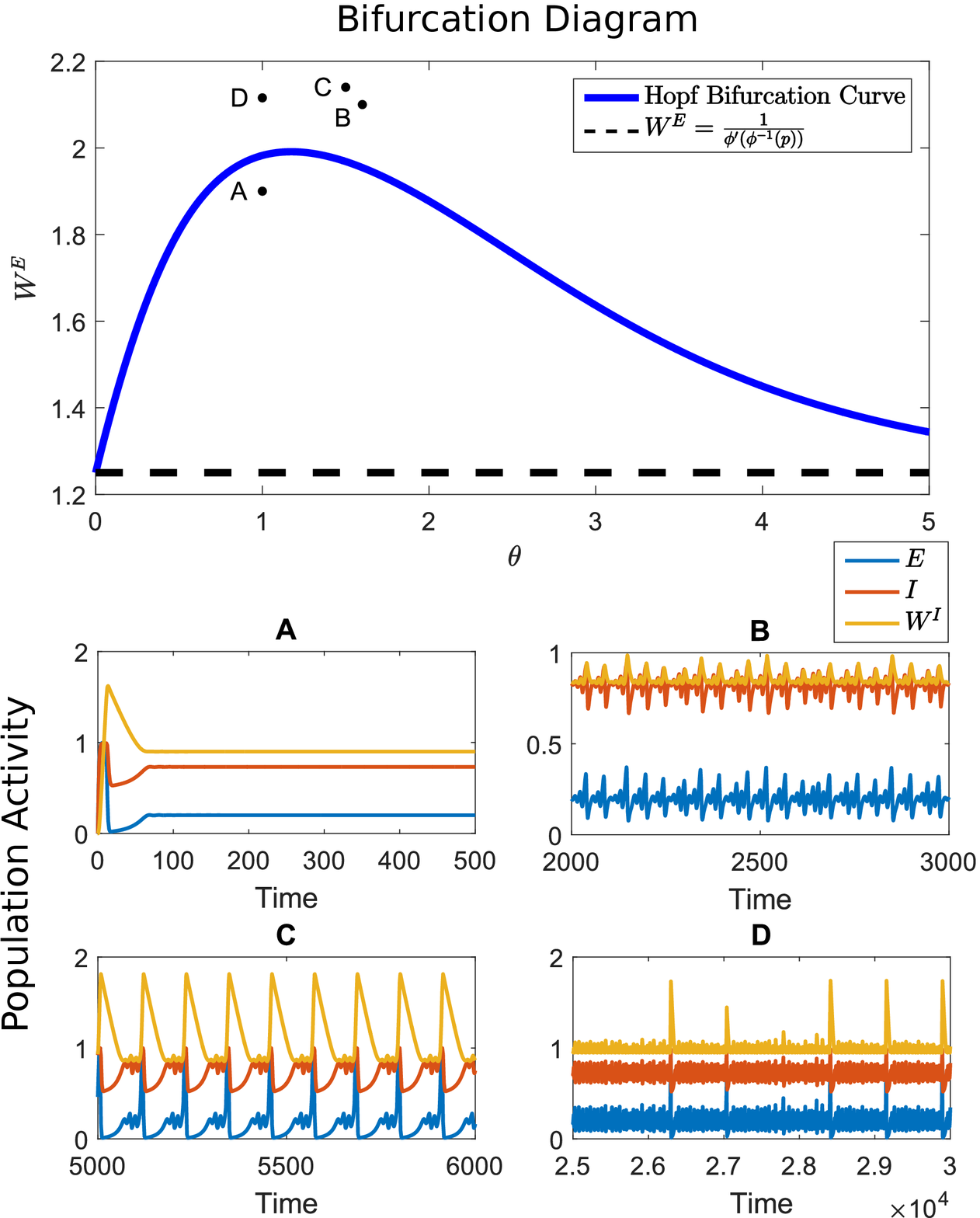}
\caption{}   \label{FIG1}
\end{figure}

For larger values of $W^E$, the system displays chaotic activity which was verified by computing the maximum Lyapunov exponent numerically (Figure \ref{FIG1}B).     This chaotic attractor contains small excursions from $\bar{\bm x}$.   Again, in this region the homeostatic mechanism is still operating within some degree of tolerance as the chaotic attractor is contained within small neighbourhood of the equilibrium.  Mixed mode oscillations are also present past the Hopf-bifurcation (Figure \ref{FIG1}C).  Surprisingly, for large enough values of $W^E$, the chaotic attractor can also contains components that operate on two separate time scales (Figure \ref{FIG1}D).  This is referred to as ``mixed mode chaos"  \cite{MMOreview,Koper}

Given the exotic nature of the mixed mode-chaos in this system, we investigated how chaos emerges in this system.  First, we fixed $\theta$ and steadily increased $W^E$ and observed a classical period doubling cascade (Figure \ref{FIG2}A,\ref{FIG2}B) to chaos.  Numerically computing the maximal Lyapunov exponent (\cite{sprott}) over the two parameter $(\theta,W^E)$ region reveals a contiguous region of chaotic solutions above the Hopf bifurcation curve (Figure \ref{FIG2}C). 

For smaller values of $W^E>W^E_{Hopf}$, the chaotic solutions are classical in nature.  For example, by plotting the $k$th maxima of the $E$ variable, $E^*_k$ as a function of $E^*_{k-1}$, we find a  stereotypical unimodal peak-to-peak or tent map \cite{Lorenz,Strogatz}.  However, as we increase $W^E$ further, a pseudo-singularity or ``pinch" emerges in the tent map at the location of the former maximum.  This is not a true singularity of this map as the set $E\in(0,1)$ is invariant.  The emergence of this singularity in the tent map corresponds to the emergence of mixed-mode chaos.  Mixed mode chaos however occurs over a narrower parameter regime for the single node.  For larger values of $W^E\gg W^E_{Hopf}(\theta)$, the system only displays large relaxation limit cycle solutions.

\subsection{Canards and Mixed Mode Oscillations} 

Next, we investigated how mixed-mode oscillations emerge in the three-dimensional, single-node case.  In particular, recent analytical work has demonstrated several cases through which long and short time scale oscillations can emerge in a three-dimensional system exhibiting a separation of time scales.  Examples include the existence folded-node case involving one fast variable and two slow variables, or the ``tourbillion" case involving two fast variables and one slow dynamical variable \cite{foldednode,MMOreview}. Both systems give rise to mixed-mode oscillations however through different mechanisms.  

We hypothesized that the most likely mechanism for the emergence of mixed-mode oscillations for our network equations was the so called tourbillion case \cite{MMOreview}.  This is due to the presence of two fast variables ($E,I$) in addition to the slow weight $W^I$.  However, the mixed-mode oscillations cannot arise from the tourbillion case in our system.  Indeed, this requires that the fast variables, given by:
\begin{eqnarray}
\tau_1 E' &=& -E + \phi\left(W^E E - W^I I \right)\label{cn1}\\
I' &=& - I + \phi\left(\theta E \right) \label{cn2}
\end{eqnarray} 
undergo a Hopf bifurcation \cite{MMOreview}.  This is not possible in any parameter set for any potential equilibrium of the system (\ref{cn1})-(\ref{cn2}) as the system is incapable of having complex eigenvalues.   In particular, the requirement for complex eigenvalues is $\frac{1}{4}\text{tr}(J)^2-\text{det}(J)<0$.  However after evaluating and simplifying this condition for (\ref{cn1})-(\ref{cn2}), we arrive at:
\begin{eqnarray}
\left(1 -\frac{1+W^E \phi'(W^E -W^I I)}{\tau_E}\right)^2 +\frac{ W^I\theta \phi'(\theta E)\phi'(W^E - W^II) }{\tau_E}>0 \label{con1}
\end{eqnarray}
and thus no complex eigenvalues for any equilibria are possible.   This immediately implies that no Hopf-bifurcation is possible and eliminates the tourbillion case from consideration.  This is a striking result as this implies that the oscillatory dynamics that emerge in the network are due to homeostasis.

With the tourbillion case removed as a possible cause of mixed-mode oscillations, we are left with several other possibilities. The time scales in our network are given by $\tau_E/\tau_I = 1$, $\tau_W/\tau_I = 5$.  Mixed mode oscillations arising from a folded-node occur when the system has one fast variable and two slow variables.  An alternate hypothesis to the tourbillion case (two fast variables and one slow variable) is that the mixed-mode oscillations occur due to the folded-node where the folded-node singularity arises when $\tau_E/\tau_I \ll 1$,  $\tau_W/\tau_I = 5$ and the qualitative features of the phase portrait merely persist until $\tau_E/\tau_I =  1$ for (\ref{WC1})-(\ref{WC3}). This latter case is the parameter range for our networks.  The folded-node case as analyzed in \cite{foldednode} demonstrates canards for the subsystem consisting of one-fast and one-slow variable, and a folded null-surface in the three-dimensions with two attracting branches and one repelling branch.  Indeed, we find a similar result in our system (Figure \ref{FIG3}).  Canards exist over a exponentially small parameter regime in the reduced $E,W^I$ system with $I$ either set to $\phi(\theta E)$ (Figure \ref{FIG3}A) or $I = \frac{1}{2}$ (not shown).   Finally, the null-surfaces for the excitatory ``fast'' variable are folded and contain two attracting regions and a repelling region.  The dynamics for large amplitude oscillations follow the attracting components of the null-surface.   This numerical analysis suggest that the mixed-mode oscillations arise from the folded-node mechanism.  Interestingly, the prototypical folded-node system also contains mixed-mode chaos when higher order terms are included in the normal-form \cite{MMOreview}.

\section{The Dual-Node Case:  Synchronous Solutions to the Single Node} \label{secDN}

As the large network equations contain no self coupling in the $\bm{W}^{EE}$ weight matrix ($\text{diag}(\bm{W^{EE}}) = \bm 0$), the single-node analysis that we have conducted is not necessarily informative of the large network dynamics.  Thus, analysis must be conducted on the simplest possible system without self-coupling, the dual-node reciprocally coupled system given by equations (\ref{dn1})-(\ref{dn6}).  In this system, the local homeostatic mechanism attempts to stabilize the excitatory activity while the opposing node functions to stimulate its neighbour.  

First, we conducted numerical simulations of the dual-node system to determine what dynamical behaviors are possible.  Surprisingly, we found that over all parameter regimes tested, the dual-node system without self-coupling synchronizes to solutions of the single-node, recurrently coupled system (Figure \ref{FIG4}A,B).  For example, the dual-node system asymptotically tends towards the same chaotic attractors, limit cycles, and mixed mode solutions as the single node system (Figure \ref{FIG4}A). For oscillatory solutions, this is not surprising as a simple derivation shows that any stable solution of the single-node system potentially corresponds to a synchronous solution in the dual-node system.   For chaotic attractors, the dual-node system exhibits a case of synchronized chaos when the parameters for both nodes are identical \cite{synchronizedchaos}.

\subsection{Local Stability Analysis of Equilibria} 

Again, due to the homeostatic nature of the dual-node system, the only equilibrium that exists is given by equation 
\begin{eqnarray} \overline{E}_1 =  \overline{E}_2 = p, \quad \overline{I}_1=\overline{I}_2= \phi( \theta p), \quad  \overline{W^I}_1=  \overline{W^I} _2= \frac{W^Ep - \phi^{-1}(p)}{\phi(\theta p )}.
\end{eqnarray}  
Furthermore, using the Jacobian to solve for the characteristic polynomial yields the following:
\begin{eqnarray}
C_{DN}(\lambda)&=&C_{SN}(\lambda)Q(\lambda)\\
Q(\lambda) &=& \lambda^3 + \lambda^2\left(\frac{1+W^E\phi'(\phi^{-1}(p))}{\tau_1}+1 \right) +  \lambda\left(\frac{1+W^E\phi'(\phi^{-1}(p))}{\tau_1} +\frac{ \overline{W^I} \phi'(\phi^{-1}(p))\phi'(\theta p)\theta}{\tau_1}+\frac{\overline{I}^2 \phi'(\phi^{-1}(p))}{\tau_1\tau_2}\right)\nonumber\\&+&\frac{\overline{I}^2 \phi'(\phi^{-1}(p))}{\tau_1\tau_2} \label{QL}.
\end{eqnarray}
Thus, instability in $C_{SN}(\lambda)$ implies instability in the dual-node system for any equilibria.  Furthermore, by the Routh-Hurwitz criterion \cite{wiggins}, all roots of $Q(\lambda)$ lie in the left complex plane if: 
\begin{eqnarray*}
\left(\frac{1+W^E\phi'(\phi^{-1}(p))}{\tau_1}+1 \right)\left(\frac{1+W^E\phi'(\phi^{-1}(p))}{\tau_1} +\frac{ \overline{W^I} \phi'(\phi^{-1}(p))\phi'(\theta p)\theta}{\tau_1}+\frac{\overline{I}^2 \phi'(\phi^{-1}(p))}{\tau_1\tau_2}\right)>\frac{\overline{I}^2 \phi'(\phi^{-1}(p))}{\tau_1\tau_2}\label{RH1}.
\end{eqnarray*}
For all physical solutions, the inequality is satisfied as the term $\frac{\bar{I}^2 \phi'(\phi^{-1}(p))}{\tau_1\tau_2}$ can be subtracted from both sides of (\ref{RH1}) with all the remaining terms on the right hand side being strictly positive.  However, roots on the right-complex plane may occur for non-physical values of these coefficients, for example when the weights are  negative.  This analysis implies that for all permissible (physical) equilibria, the stability of the dual node system is directly inherited from the stability of the single node.    In other words, for symmetrically coupled systems, the local homeostasis rule loses no robustness in regulating network dynamics up to the Hopf-bifurcation and the recurrent inhibition can counteract non-local excitation just as well as local excitation.    
 
\subsection{Stability Analysis of Limit Cycles} 
If we consider any limit cycle for the single node-system $\bm x(t) = (E(t),I(t),W^I(t))$ such that $\bm x(t) = \bm x(t+T)$ for some $T>0$, $\forall t$, then the following is an admissable limit cycle solution to the dual-node system:
\begin{eqnarray}
\bm z(t) = \begin{pmatrix}\bm x(t)\\ \bm x(t) \end{pmatrix}
\end{eqnarray}
with an identical period $T$, for all $t>0$.  Furthermore, if we consider the monodromy matrix system
\begin{eqnarray}
\dot{\bm \epsilon} = A(\bm x(t)) \bm \epsilon
\end{eqnarray}
derived by linearizing equations (\ref{WC1})-(\ref{WC3}) around $\bm x(t)$ then linearization for equations (\ref{dn1})-(\ref{dn6}) 
can be written as 
\begin{eqnarray}
\dot{\bm \epsilon} &=& A(\bm x(t))\bm\epsilon +W^E\phi'(W^EE(t) - I(t)W^I(t))(\nu_1(t)-\epsilon_1(t))\label{vareq1}\\
 \dot{\bm \nu} &=& A(\bm x(t))\bm\nu +W^E\phi'(W^EE(t) - I(t)W^I(t))(\epsilon_{1}(t) - \nu_1(t))\label{vareq2}.
\end{eqnarray}
In order to analyze the stability of limit cycles, we require a fundamental solution set to equations (\ref{vareq1})-(\ref{vareq2}).  First, if we consider $\bm \Upsilon (t) = [\bm \epsilon_1(t),\bm \epsilon_2(t) \bm \epsilon_3(t)]$, then three fundamental solutions are immediately given by $[\epsilon_k(t),\epsilon_k(t)]$ for $k=1,2,3$.  This implies that if the limit cycle is unstable in the single node system, (\ref{WC1})-(\ref{WC3}), then it is unstable in the dual node system.  We leave the stability analysis of these limit cycles and other trajectories for future work.

\section{The Fully Coupled $N$-Node System} \label{secFN}

As we have previously demonstrated, the dual-node system without self coupling has identical dynamics to the single-node, self coupled system and even exhibits chaotic synchronization to identical attractors as the single-node.  Thus, the single node is largely predictive of the qualitative dynamics of the coupled system despite the removal of self-coupling.   Thus, we investigated if a similar result would apply to the large uncoupled system given by equations (\ref{fn1})-(\ref{fn3}).    First we analyzed a pair of analytically resolvable cases for matrices that satisfied specific assumptions.  Then, we numerically explored the system (\ref{fn1})-(\ref{fn3}) coupled by the weight matrix considered in \cite{Peter}.  

\subsection{Exactly Resolvable Cases} \label{ana1}

First, we considered a pair of analytically resolvable cases.   
If we consider the all-to-all coupled matrix:
\begin{eqnarray}
\bm{W^{EE}}_{ij} =\begin{cases}  \frac{W^E}{N_E-1} & i\neq j \\ 0 & i=j \end{cases} \label{MF1}
\end{eqnarray}
then the characteristic polynomial reduces to:
\begin{eqnarray}
C_{N_E}(\lambda) = \hat{Q}(\lambda)^{N_E - 1} C_{SN}(\lambda) 
\end{eqnarray}
where $\hat{Q}(\lambda)$ and $C_{SN}(\lambda)$ are the polynomials resolved in the dual node (equation (\ref{QL})) and single node case  (See Supplementary Materials Section S2).   The polynomial $\hat{Q}(\lambda)$ however has $\bar{W^E}$ in place of $W^E$.  Our previous analysis immediately applies and shows that with the coupling matrix (\ref{MF1}), the system (\ref{fn1})-(\ref{fn3}) has identical local stability to the single node.  We refer to equation (\ref{MF1}) as the ``mean-field'' assumption.  Self-coupling need not be removed in this case, however the results will differ slightly from the single node if self-coupling is considered.  

Finally, we remark that there is one other case where the stability of the system can be determined analytically, when the row sum of the coupling weight matrix is constant:
\begin{eqnarray}
\sum_{j=1}^{N_E} \bm W^{EE}_{ij} = W^E, \quad i = 1,2,\ldots N_E
\end{eqnarray}

We decompose the weight matrix $\bm{W^{EE}} = W^E \cdot\bm L^{EE}$ where the row sum of $\bm L^{EE}$ is equal to one.   The scalar term $W^E$ scales the magnitude of the components of the weight matrix, similar to $W^E$ in the single and dual node cases.   For this case, one can resolve the eigenvalue spectrum explicitly as the characteristic polynomial factors readily:
\begin{eqnarray}
C(\lambda) &=& \prod_{i=1}^{N_E} \left(\tilde{Q}(\lambda) -r_i  \frac{ \lambda(\lambda+1)\phi'(\phi^{-1}(p))}{\tau_1}W^E  \right)
\end{eqnarray}
where each $r_i$ is an eigenvalue of the weight matrix $\bm L^{EE}$. Each $\tilde{Q}(\lambda)$ is a cubic polynomial given by:
\begin{eqnarray}
\hat{Q}(\lambda) &=& \lambda^3 + \lambda^2\left(\frac{1}{\tau_1}+1 \right) +  \lambda\left(\frac{1}{\tau_1} +\frac{ \overline{W^I} \phi'(\phi^{-1}(p))\phi'(\theta p)\theta}{\tau_1}+\frac{\overline{I}^2 \phi'(\phi^{-1}(p))}{\tau_1\tau_2}\right)\nonumber+\frac{\overline{I}^2 \phi'(\phi^{-1}(p))}{\tau_1\tau_2}. \label{QL2}\\
\end{eqnarray}
   The steady states $\bar{W^I}$ and $\bar{I}$ are given by identical formulas as in the single and dual node cases.   Given the structure of the polynomial $Q(\lambda)$, this yields a Hopf-bifurcation immediately through an identical derivation in the single node case.   The curve will be of the form:
\begin{eqnarray}
W^E_{Hopf,i}(\theta) &=& \frac{1}{r_i\phi'(\phi^{-1}(p))}\left(1-\tau_1 \mu_+(\theta)\right) \label{Hopf3}\\
r_i &=& \max_{i=1\ldots N_E}\{r_i\}
\end{eqnarray}
where $\mu_+(\theta)$ is redefined and $r_i$ is an eigenvalue of $\bm W^{EE}$ (see Supplementary Materials).   As $W^E$ is increased, the first intersection of $W^E = W^E_{Hopf,i}$ determines the Hopf bifurcation curve.  For $\theta\gg1$ and $ \theta \ll 1$, this is readily seen to be the curve corresponding to the largest positive eigenvalue of $\bm L^{EE}$.   

  Additionally, if the row-sum of the matrix $\bm W^{EE}$ is non-constant, but narrowly distributed, one can still approximate the Hopf-bifurcation curve by using the mean-row sum (see Supplementary Materials).  We validate this approximation in the subsequent section as applied to the weight matrix considered in \cite{Peter}.  

\subsection{Numerical Exploration of the Experimentally Coupled System} 
  The connectivity matrix, $\bm W^{EE} =  W\bm L^{EE}$, is derived from functional neuroimaging data and is described in greater detail in \cite{Peter,Honey,hagmann}.  The matrix $\bm L^{EE}$ is shown in Figure \ref{FIG5}A.  The matrix couples 66 homesotatically regulated Wilson-Cowan nodes.  Furthermore, $\bm L^{EE}_{ii} = 0$ for all $i$ and thus the nodes contain no self-coupling. As our single and dual node analyses indicate a branch of Hopf bifurcations, we numerically computed the eigenvalues over the two parameter $(W,\theta)$ space and searched for the first eigenvalue $\lambda_i$ crossing $\text{Re}(\lambda_i)=0$ as a function of $\theta$ for each value of $W$.  This yielded a similar potential Hopf-bifurcation curve as the single and dual node cases.  The curve was again unimodal with identical asymptotes as $\theta \rightarrow 0$ and $\theta \rightarrow \infty$.  We conducted large scale numerical simulations to determine if the curve indeed indicated a transition from steady state dynamics to oscillations.  For $W<W_{Hopf}(\theta)$, we observe decay to a steady state equilibrium and oscillations or chaos for $W>W_{Hopf}(\theta)$ (Figure \ref{FIG5}B,\ref{FIG5}C).   Finally, we applied the analytical approximation derived in section \ref{ana1} for comparison.  The approximation has the greatest accuracy near the asymptotes $\theta \rightarrow 0$ and $\theta \rightarrow \infty$ and indicates that the common asymptotic behavior for $W_{Hopf}$ is:
  \begin{eqnarray}
  W_{Hopf}(\theta)\sim \frac{1}{\phi'(\phi^{-1}(p))r^{max}_i}, \quad \theta \rightarrow \infty, \theta \rightarrow 0
  \end{eqnarray}
  where $r^{max}_i$ is the large positive eigenvalue of $\bm L^{EE}$.  As in our analysis of the single node, this asymptotic behavior corresponds to the region of guaranteed stability of the steady state for $W^E < W^E_{Hopf}(0)$

As in the single and dual node cases, the large network also displays mixed mode oscillations and mixed mode chaos (Figure \ref{FIG5}C,D).  Interestingly, due to the  heterogeneous coupling in the weight matrix, the nodes do not all transition to chaotic dynamics in an identical fashion (Figure  \ref{FIG5}E).  This is despite the connectivity in the network only being moderately sparse ($p=0.2635$). For example, some nodes can display a smaller attractor without mixed mode elements, other nodes contain larger amplitude components while others are essentially still stabilized around their equilibrium point with minimal interference from the rest of the network.  Also note that the attractors in Figure \ref{FIG5}F occupy a similar region of the reduced phase space $(E,I,W^I)$ as the single and dual node cases when we plot every node $(\bm E_k, \bm I_k, \bm W^I_k)$ in the same reduced phase space.  

Given the heterogeneity in the chaotic dynamics of the individual nodes in the coupled networks, we investigated whether node and connection deletion might enhance the stability of the homeostatic mechanism.  Indeed, the homeostatic mechanism is inherently local and trying to stabilize the dynamics of the individual nodes despite receiving external, potentially destabilizing inputs.  To that end, we deleted a node and recomputed our Hopf bifurcation curves for each node deletion yielding 66 different systems with 65 nodes.  In every system, the deletion either had minimal effect on the Hopf-bifurcation curve or it shifted the curve upward.   Thus, deleting either connections or nodes can only increase the stability of the homeostatically induced equilibrium.  The maximum change was a 14.69\% shift upwards (as measured from the peak) given by deleting the 25th node.  Deleting individuals connections in the weight matrix, $\bm W^{EE}$ yielded at most a 4.01\% shift upwards in the Hopf bifurcation curve.  Interestingly, the largest shift in the Hopf-bifurcation curve corresponds to $\bm{W}^{EE}_{21,4}$ and not node 25.   Further work is required to determine if there are any properties of this node to the weight matrix that grants it a disproportional effect on the network dynamics.

\section{Discussion} 

Through a combination of numerical and analytical work, we studied a homeostatically regulated Wilson-Cowan system in three separate cases: isolated single-nodes, reciprocally coupled dual-nodes, and large coupled networks where the connection strength was derived from functional neuroimaging data \cite{Peter,Honey,hagmann}.   We found that the isolated single node displays a plethora of complex dynamics such as mixed mode oscillations, chaos via a period-doubling cascade, and mixed-mode chaos.   The source of these rich dynamics is the folded node recently analyzed in \cite{foldednode}.  Two nodes with no self coupling and symmetric reciprocal excitatory coupling acted essentially as a single, self-coupled node and synchronized to the steady state attractors in the single node-case.  We demonstrated analytically that the stability of steady states in the single node case is directly inherited in the dual node case.  Furthermore, any unstable limit cycle in the single node is unstable in the dual node case.  Finally, we numerically explored the large coupled network and showed a similar transition to oscillatory behavior for strong enough excitatory coupling.  The individual nodes in the large network displayed similar dynamics to isolated recurrently coupled nodes in different parameter regimes.  Interestingly, node deletion and connection deletion yielded non-trivial increases in the stability of the homeostatic set point for all values of excitatory to inhibitory coupling.   

Past the Hopf-bifurcation, the network exhibits a rich dynamical repertoire consisting of oscillatory activity, chaos, and mixed-mode elements of both.  Whether these dynamical states are potentially functional or pathological remains to be seen.  Indeed, even for the experimentally determined chaotic attractors in \cite{destexhe}, some correspond to functional states such as stages of sleep while others correspond to pathological states such as epileptic seizures.  In the former case, we have demonstrated that synaptic homeostasis can support the emergence of complex dynamics.  If however, these states are pathological, then they represent a failure of homeostasis in regulating network dynamics.  Our node-deletion and connection deletion experiments demonstrate that the deletion of even single nodes or connections can increase the stability of the entire network through a shift in the Hopf-bifurcation curve upwards.

Homeostasis is widely regarded as a mechanism for the maintenance of network dynamics, and more specifically the maintenance of a steady-state average firing rate \cite{homeostasis1,homeostasis2,homeostasis3} and is regarded as a stabilizing force in network dynamics \cite{homeostasis4}.  This steady-state is regulated at slow time scales on the order of minutes \cite{homeostasis2} or hours \cite{homeostasis5}.  For example, the homeostatic model in \cite{vogels} was shown to maintain the asynchronous irregular regime where neurons fire irregularly, but at a constant average rate.  It is thus surprising that low dimensional yet rich structures such as mixed-mode chaotic attractors emerge under the presence of homeostasis.   

While mixed-mode chaos is a relatively understudied phenomenon, it has been previously documented in the literature.  For example, the following polynomial system: 
\begin{eqnarray}
\epsilon \dot{x} &=& y - x^2 - x^3 \\
\dot{y} &=& z-x\\
\dot{z} &=& -\nu - ax - by -c z 
\end{eqnarray}
also exhibits mixed-mode chaos, as demonstrate in \cite{MMOreview}.  This model is also related to an earlier chemical reaction model in \cite{Koper} and was also derived as a generic reduction of a mixed-mode oscillatory system in \cite{krupa}.  Mixed-mode chaos was also observed in \cite{MMCHAOS1}.  The authors analyze an enzymatic reaction scheme and show similar pinched/singular tent-maps for the mixed-mode chaotic attractors they observe.  Interestingly, the authors suggest a homoclinic limit cycle as their return mechanism.  The authors suggest a version of the classical Shilnikov bifurcation resulting in homoclinic chaos \cite{kuznetsov}.

Our results demonstrate that the rich dynamical states demonstrated in \cite{Peter} are an intrinsic property of synaptic homeostasis, which is capable of more than stabilizing average firing rates across a network.  With inhibitory synaptic homeostasis, stability can only be guaranteed up to a point in the parameter space.  This point  is analytically determined and is related to the properties of the tuning curves, the homeostatic set point, and the connectivity between excitatory populations.
  The resulting dynamics past this point displaying a rich dynamical repertoire including oscillations and chaos, both of which can occur on two different time scales.   This is an intrinsic consequence of the inhibitory synaptic homeostasis rule as the two-dimensional Wilson-Cowan node that we consider is incapable of oscillating without inhibitory synaptic homeostasis.   These dynamical repertoires might have functional or pathological consequences for populations of neurons.  
\newpage 

\begin{table}
\begin{center}
\begin{tabular}{ |c|c| } 
 \hline
 Parameter  & Numerical Value\\
 \hline
 $a$ & 5 \\ 
  \hline
 $p$  & 0.2  \\ 
  \hline
 $\tau_E$  & 1 \\ 
  \hline
 $\tau_I$ & 1 \\ 
  \hline
 $\tau_W$  & 5  \\ 
 \hline
 $W^E$ & (see Figure Captions, typically [0,3] \\
 \hline
 $\theta$& (see Figure Captions, typically [0,10]\\
 \hline
\end{tabular}
\end{center}
\caption{The parameter values for the system of equations (\ref{WC1})-(\ref{WC3}) (single node), (\ref{dn1})-(\ref{dn6}) (dual node), and (\ref{fn1})-(\ref{fn3}) (full network).  Note that for the full network equations, $\bm W^{IE} =\theta\bm I_{N}$, where $\bm I_N$ is the $N$ dimensional identity matrix, and $N$ consists of the number of nodes. }\label{TAB1}
\end{table}

\newpage
\section*{Figures}

\newpage

\begin{figure}[htp!]
\centering
\includegraphics[scale=0.45]{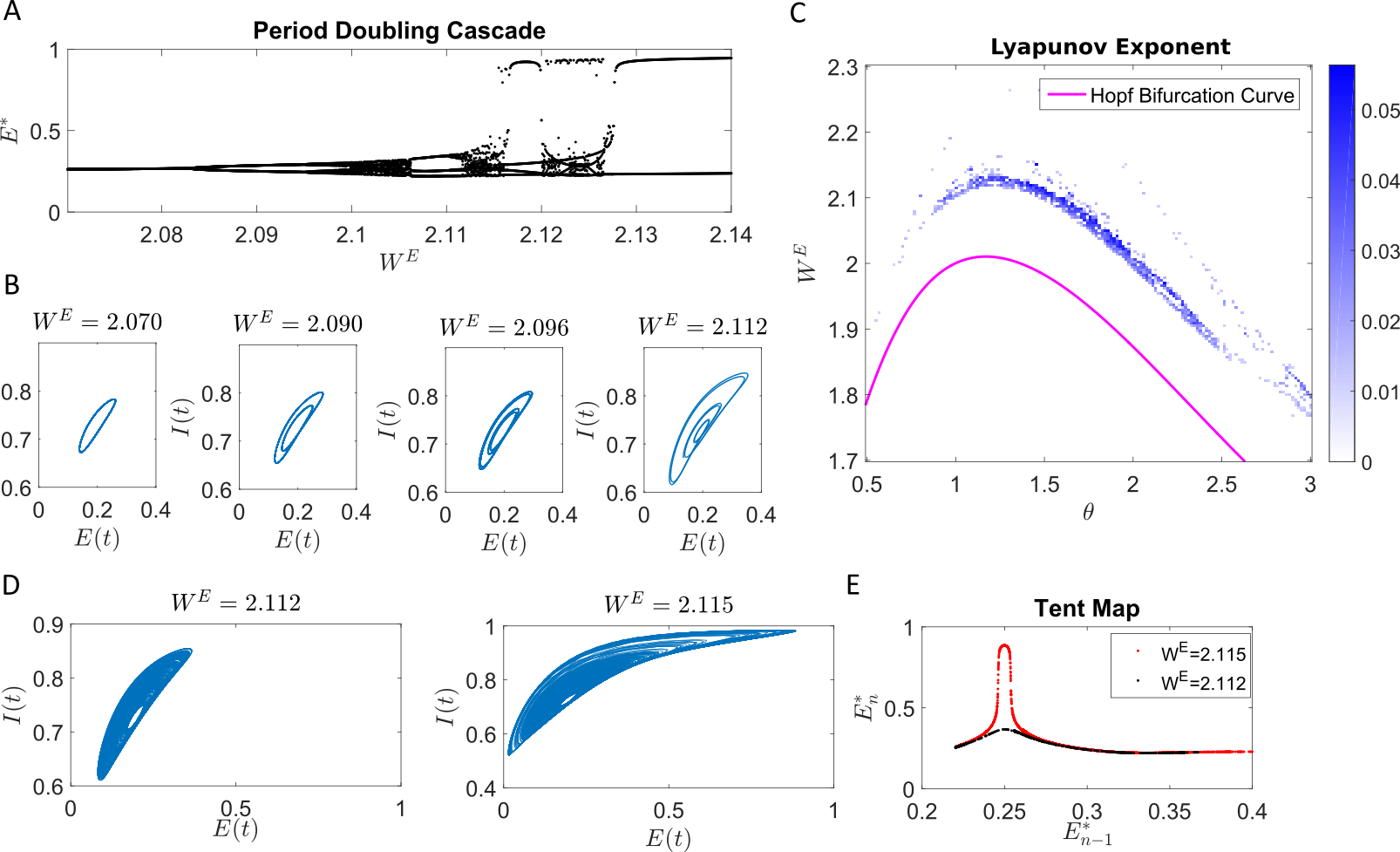}
\caption{ }   \label{FIG2}
\end{figure}
\newpage

\begin{figure}[htp!]
\centering
\includegraphics[scale=0.65]{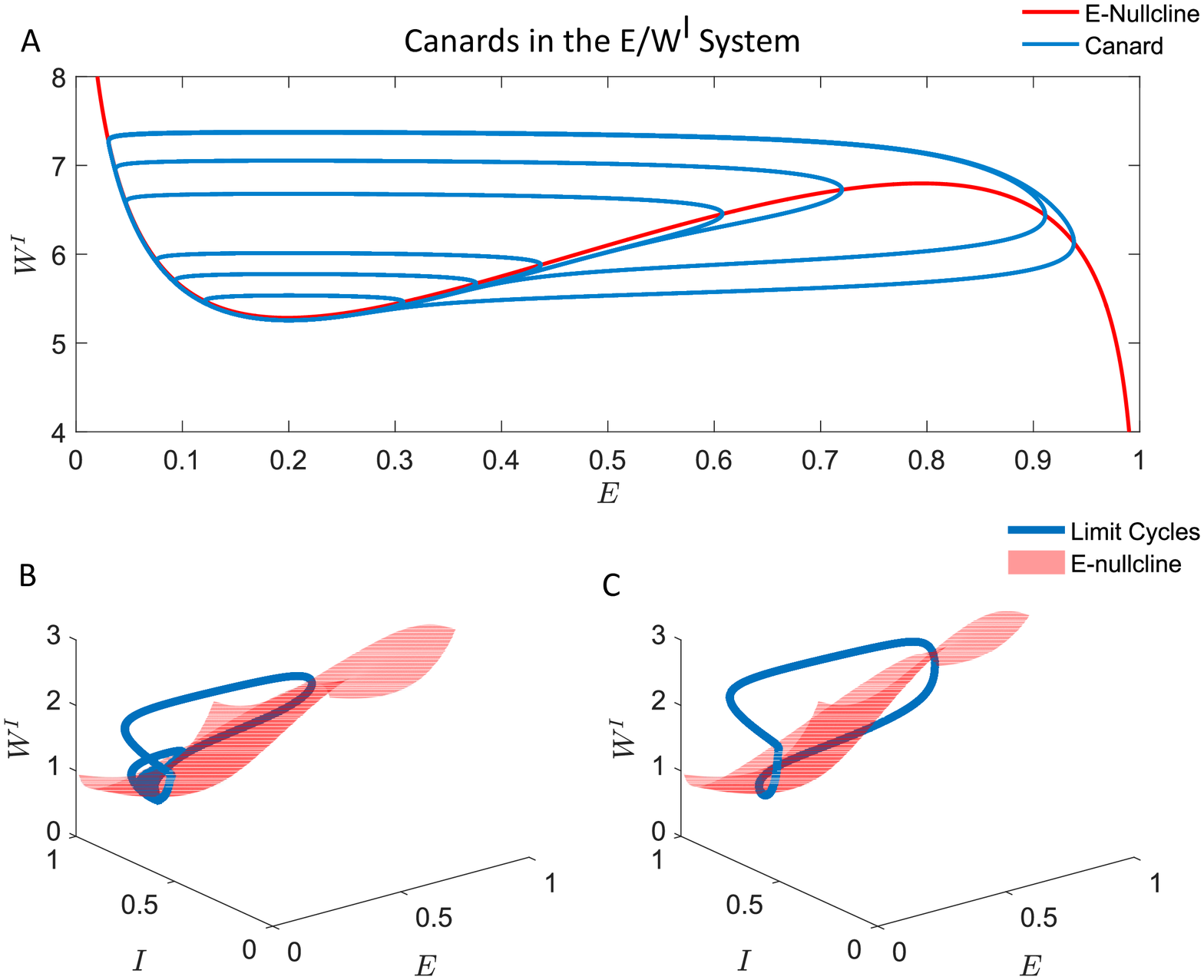}
\caption{ }   \label{FIG3}
\end{figure}
\newpage

\begin{figure}[htp!]
\centering
\includegraphics[scale=0.5]{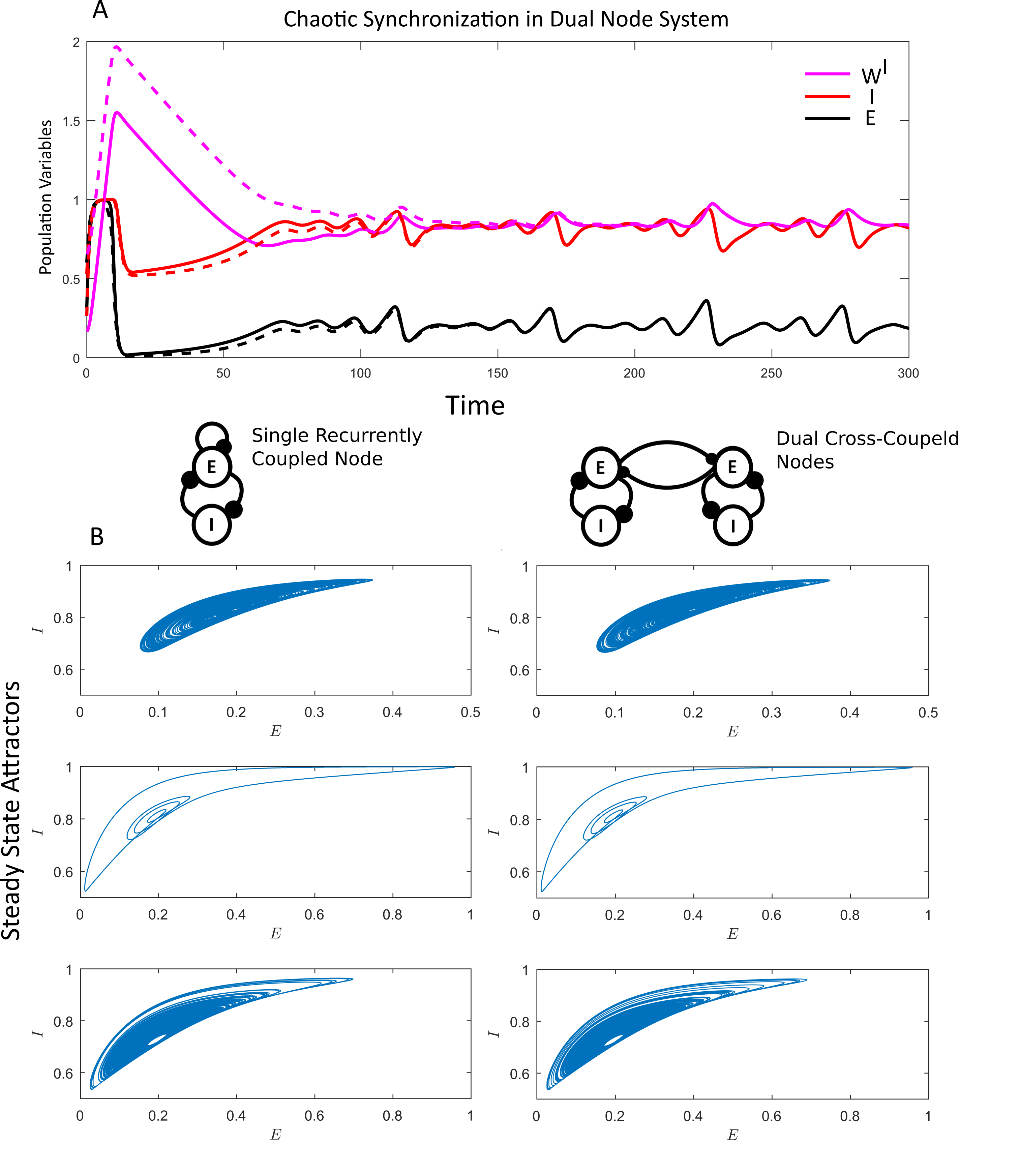}
\caption{}    \label{FIG4}
\end{figure}

\begin{figure}[htp!]
\centering
\includegraphics[scale=0.55]{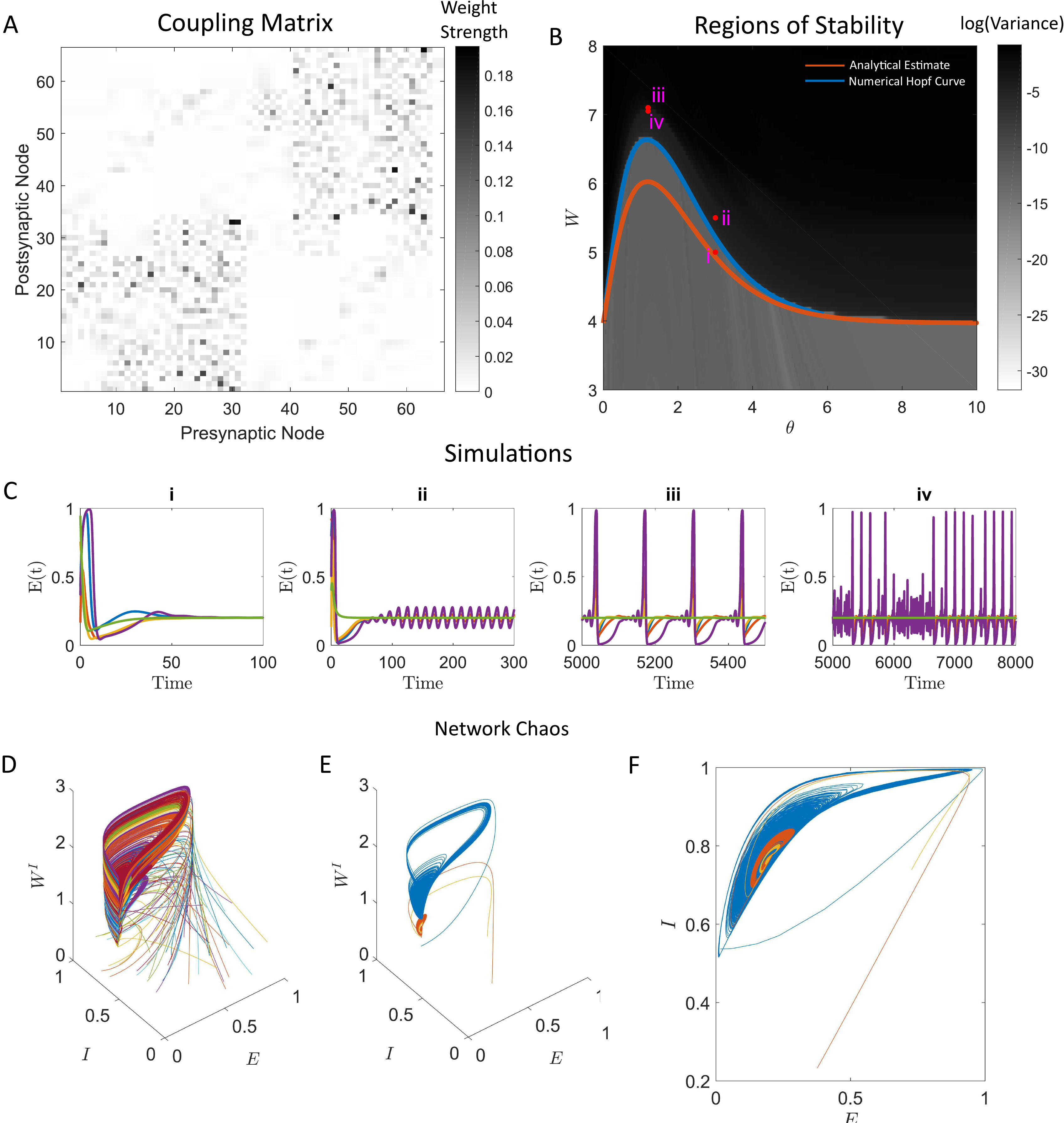}
\caption{}    \label{FIG5}
\end{figure}
\newpage

\section*{Figure Captions}

\subsection*{Figure 1}

(Top) The Hopf bifurcation curve for the single-node system can be derived explicitly. Analysis of the $\sigma = 0$ case coupled with numerics demonstrates that the bifurcation is a supercritical bifurcation.  As we vary the $(\theta,W^E)$ parameters, different behaviors emerge corresponding to (A) stability of the target activity, (B) chaotic loss of stability (C) mixed mode oscillations, and (D) mixed mode chaos.  The parameters were $p=0.2$, $\tau_1 = 1$, $\tau_2 = 5$ with $(\theta,W^E)$: (1,1.9), (1.6,2.1), (1.5,2.14), (1,2.115) for (A)-(D), respectively.  All simulations were conducted in MATLAB \cite{MATLAB} using the ode45 integration suite to implement a Runge-Kutta 4th order integration scheme.  

\subsection*{Figure 2}

(A) The maxima of limit cycles are plotted as a function of the recurrent self coupling, $W^E$ for the single node system.  As $W^E$ increases past $W^E_{Hopf}(\theta)$, a period doubling cascade to chaos occurs.  (B) The limit cycles and chaotic attractor plotted for increasing values of $W^E$.  (C) The maximum Lyapunov exponent is computed over the two parameter $(\theta,W^E)$ region showing patches of chaos that onset after the Hopf bifurcation curve.  (D) The chaotic attractor for sub-threshold and mixed mode chaotic solutions.  (E) As $W^E$ is increased past $W^E_{Hopf}(\theta)$, the period doubling cascade produces a tent map similar to the classical Lorenz tent map.  For larger values of $W^E$ the tent map develops a pseudo-singularity at the maximum value.  Note that this is not strictly a singularity in the tent map as the dynamics of the $E$ are restricted to $E\in(0,1)$.  For all simulations in (A),(B), and (E), $\theta=1$ was used.

\subsection*{Figure 3} 
(A) Canard limit cycles for the dual node system given by equations (\ref{cn1})-(\ref{cn2}).  The seven limit cycles show a rapid increase in amplitude shortly after a supercritical Hopf bifurcation.  The $W^E$ parameter for 6 limit cycles agrees to four decimal places ($W^E=7.5959$).  The final limit cycle is a large relaxation limit cycle ($W^E=7.6$).  The limit cycles were computed with direct simulation of the ordinary differential equations (\ref{cn1})-(\ref{cn2}) using a (4,5) order Runge-Kutta scheme.  The $\theta$ parameter was fixed at $\theta = 1$.  (B)  Shown above is the period doubled limit cycle (teal) for the system (\ref{WC1})-(\ref{WC3}) in addition to the $I$-nullcline (blue).  Under the assumption that both the inhibition and the homeostatic mechanism are operating as fast variables, we can see the mixed-mode oscillations arising from the so-called ``folded-node case".  The $(\theta,W^E)$ parameters were $(2,2.02)$   (C) A relaxation cycle emerges with increasing values of $W^E$.  For all simulations, $\theta =1$ was fixed.  The $(\theta,W^E)$ parameters were $(2.5,2)$

\subsection*{Figure 4}

(A) Shown above is the time series for the symmetrically coupled dual node system without self-coupling.  The nodes synchronize with each other to a solution state for the single node system at steady-state, independent of where in the parameter region we are or the characteristics of the steady state.  The first node is showed in sold lines with the excitation (black), inhibition (red), and homeostatic weight (magenta).  The second node is plotted as a dashed line.   The parameter set in the $(W,\theta)$ space are $(1.6,2.1)$.  (B) The steady state attractors for the single node (left) and the dual node (right) are plotted in the $(E,I)$ projection space.  The parameters in the $(W,\theta)$ space (1.6,2.1) (top) which corresponds to a chaotic attractor, $(1.5,2.14)$ (middle) which corresponds to a mixed-mode oscillation, and $(1,2.115)$ (bottom) which corresponds to mixed-mode chaos.  Note that in all cases, the steady state attractors are identical for either the single recurrently coupled node or the dual-node symmetrically coupled nodes.   Only one node is plotted in the dual-node case, however due to synchrony, the trajectory for the second node is identical.   

\subsection*{Figure 5}

(A) The coupling matrix used to connect the excitatory components of the nodes.  Note that the matrix is highly structured, and contains no elements on its diagonal (no self-coupling).  The system consists of 66 nodes.  (B) The Hopf bifurcation (blue) curve is determined manually by evaluating the eigenvalues over the two parameter ($\theta,W$) space numerically and plotting the level set for the first eigenvalue crossing $\text{Re}(\lambda_i)=0$.  This curve was verified by running a mesh of simulations over the ($\theta,W$) parameter space that consisted of $2\times 10^4$ time units each.  The final half of the simulation was used to compute the log of the variance of $E_1(t)$ to determine if the equilibrium was stable. Blacker values correspond to either limit cycles or chaos and a loss of stability as the dynamics no longer settle onto a steady state $E_1(t) =p$.   Additionally, the analytical approximation (in orange) which assumes that the row-sum of the matrix $\bm W^{EE}$ is approximately constant is also plotted.  The accuracy is highest at the asymptotes $(\theta \gg 1, \theta\ll 1)$.  
The four parameter points (i)-(iv) are shown in (C) in addition to their relationship with the Hopf-Bifurcation curve.  The parameter values in the $(W,\theta)$ plane are (5,3),(5.5,3),(7.1,1.2), and (7.05,1.2) for (i)-(iv), respectively.  Note that the Hopf-bifurcation curve has a similar shape and qualitative behavior to the curve in the single-node/dual-node case.  (C) For the parameter values shown, the large network also displays a decay to a static equilibrium for $W<W_{Hopf}(\theta)$, stable oscillations for $W>W_{Hopf}(\theta)$, mixed mode oscillations, and mixed-mode chaos.  (D) All nodes are plotted in a 3D phase portrait for the parameter region (iv) demonstrating the chaotic attractor.  (E) Three nodes are plotted from the full 66-dimensional system in the same phase space.  Some nodes in the full system display mixed-mode chaos simultaneously to other nodes that display generic chaos or very small chaotic deviations from the steady state equilibrium. (F) The same figure (E) only projected down to the $(E,I)$ phase space for comparison purposes with Figure \ref{FIG2}D.

\newpage
\bibliographystyle{apalike}	
\bibliography{chaosbib}

\newpage
\section*{Supplementary Material} 

\section*{S1:  First Lyapunov Coefficient for the $\theta = 0$ Case}

We can compute the Lyapunov coefficient for the Hopf bifurcation point when $\theta = 0$ quite easily without having to resort to the center-manifold theorem.  By setting $\theta =0$, we have the following system 
\begin{eqnarray}
\tau_1 E' &=& -E + \phi\left(W^E E - W^I I \right)\\
I' &=& -I + \phi(0) \\ 
\tau_2 {W^I}' &=& I(E-p)    
\end{eqnarray} 
and in essence $I(t)$ has become decoupled from the other equations and can be set to its equilibrium value of $\phi(0)$.  This reduction yields 
\begin{eqnarray}
\tau_1 E' &=& -E + \phi\left(W^E E -{W}^I \right)\\
\tau_2{{W}^I}' &=& (E-p)    
\end{eqnarray} 
after rescaling $\tau_2$ and $W^I$ to absorb $I=\phi(0)$.  To proceed, we shift the equilibrium to the origin
\begin{eqnarray}
\hat{E} &=& E - p \\
\hat{W} &=& W^I - W^Ep +\phi^{-1}(p)
\end{eqnarray}
which yields the following 
\begin{eqnarray*}
\tau_1 \hat{E}' &=& - \hat{E} - p + \phi( W^E(p+\hat{E}) -(\hat{W}+W^E p - \phi^{-1}(p)) \\
&=& -\hat{E} - p + \phi(W^E \hat{E}-\hat{W} + W^Ep - W^Ep + \phi^{-1}(p)) \\
&=& -\hat{E} -p+ \phi(\phi^{-1}(p)) +  {\phi'(\phi^{-1}(p))}\phi(W^E \hat{E}-\hat{W} )+ \frac{\phi''(\phi^{-1}(p))}{2!}\phi(W^E \hat{E}-\hat{W} )^2 \\
&+&  \frac{\phi'''(\phi^{-1}(p))}{3!}\phi(W^E \hat{E}-\hat{W} )^3 + H.O.T. \\
&=& -\hat{E}(1 - W^E \phi'(\phi^{-1}(p)))-\phi'(\phi^{-1}(p))\hat{W}  + \frac{\phi''(\phi^{-1}(p))}{2!}\phi(W^E \hat{E}-\hat{W} )^2 \\
&+&  \frac{\phi'''(\phi^{-1}(p))}{3!}\phi(W^E \hat{E}-\hat{W} )^3 + H.O.T. \\
\tau_2\hat{W}'&=& \hat{E} 
\end{eqnarray*}
where H.O.T. denotes Higher Order Terms.  
If we take the bifurcation condition 
$$ W^E = \phi'(\phi^{-1}p)^{-1} = \frac{1}{ap(1-p)}$$ 
we obtain the following system 
\begin{eqnarray}
\tau_1\hat{E}' &=& -{ap(1-p))}\hat{W} +  \frac{\phi''(\phi^{-1}(p))}{2!}\phi(W^E \hat{E}-\hat{W} )^2 
+ \frac{\phi'''(\phi^{-1}(p))}{3!}\phi(W^E \hat{E}-\hat{W} )^3 + H.O.T. \\
\hat{W}' &=&  \frac{\hat{E}}{\tau_2} 
\end{eqnarray}
This system can be transformed with $\hat{E} = K \tilde{E}$ into a system of the form 
\begin{eqnarray}
\hat{E} &=& -\omega \hat{W} +  \frac{\phi''(\phi^{-1}(p))}{2!K\tau_E}\phi(W^E \hat{E}-\hat{W} )^2 
+ \frac{\phi'''(\phi^{-1}(p))}{3!K\tau_E}\phi(W^E \hat{E}-\hat{W} )^3 + H.O.T. \\
\\ &=& -\omega \hat{W} + P(\hat{E},\hat{W})\\ 
\hat{W}' &=& \omega \hat{E} 
\end{eqnarray}
where 

$$ K = \sqrt{ \frac{\tau_2 ap(1-p)}{\tau_1}}, \quad \omega = \sqrt{\frac{ ap(1-p)}{\tau_2\tau_1}}$$  
which is the standard form to compute the Lyapunov Coefficient for a two-dimensional function.  The formula for the Lypaunov coefficient can be found in \cite{GUCK}  The resulting computation yields 
$$l_1(0) =-\sqrt{a}\frac{(1 + a p \frac{\tau_1}{\tau_2}(1-p) )}{(\tau_1/\tau_2)^{3/2} (1-p)\sqrt{1-p}}<0 $$ 
 and thus the bifurcation is supercritical, for $p\in(0,1)$.  

\section*{S2 Local Stability Analysis of Equilibria in $N$-node Coupled System} 
\subsection*{The Mean-Field Solution}
Here, we will consider a simple-case where the $N$ node system without self-coupling is also analytically resolvable for the Hopf-bifurcation.  In particular, consider the following conditions: 
\begin{eqnarray}
 \quad W^E_{ii} = 0,   \quad W^{E}_{ij}=\frac{W^E}{N^E-1} = \bar{W}^E \label{eq1}
\end{eqnarray} 
These solutions correspond to the mean-field of the $N_E$ nodes.  Note that the equilibria of the system (\ref{fn1})-(\ref{fn2}) are unchanged, independent of  the conditions (\ref{eq1}).  However, under the mean-field conditions (\ref{eq1}), the stability criterion are resolvable.    In particular, suppose we reorder the $3N_E$ equations such that the first $N_E$ equations correspond to $E_i$, the next $N_E$ correspond to $I_i$, and the final $N_E$ correspond to $W^I_i$.  Then the Jacobian can be written block matrix form: 
\begin{eqnarray}
J = \begin{pmatrix} -\left(\frac{1}{\tau_1} +\frac{\bar{W}^E\phi'(\phi^{-1}(p))}{\tau_1} \right)\bm I_{N_E} + \frac{\bar{W}^E \phi'(\phi^{-1}(p))}{\tau_1}  \bm 1_{N_E} & -\frac{\overline{W}^I\phi'(\phi^{-1}(p))}{\tau_1}\bm I_{N_E} &  -\frac{\overline{I}\phi'(\phi^{-1}(p))}{\tau_1}\bm I_{N_E}\\ 
\phi'(\theta p) \theta \bm I_{N_E} & - \bm I_{N_E} & \bm 0_{N_E}\\
\frac{\bar{I}}{\tau_2}\bm I_{N_E} & \bm 0 _{N_E} & \bm 0 _{N_E} 
\end{pmatrix} 
\end{eqnarray} 
where $\bm I_{N_E}$, $\bm 1_{N_E}$ and $\bm 0_{N_E}$ denote the $N_E$ identity matrix, and $N_E\times N_E$ matrices where all elements are 1 or 0, respectively.  The characteristic polynomial is given by the following:

\begin{eqnarray}
C(\lambda)  &=& \det \begin{pmatrix} -\left(\frac{1}{\tau_1} +\frac{\bar{W}^E\phi'(\phi^{-1}(p))}{\tau_1} \right)\bm I_{N_E} + \frac{\bar{W}^E\phi'(\phi^{-1}(p))}{\tau_1}  \bm 1_{N_E} -  \bm I_{N_E}\lambda  & -\frac{\overline{W}^I\phi'(\phi^{-1}(p))}{\tau_1}\bm I_{N_E} &  -\frac{\overline{I}\phi'(\phi^{-1}(p))}{\tau_1}\bm I_{N_E}\nonumber \\ 
\phi'(\theta p) \theta \bm I_{N_E} & - \bm I_{N_E}(1+\lambda) & \bm 0_{N_E}\nonumber\\
\frac{\bar{I}}{\tau_2}\bm I_{N_E} & \bm 0 _{N_E} &  - \bm I_{N_E}\lambda
\end{pmatrix}\nonumber \\
&=& \det\bigg(-\left(\left[\frac{1}{\tau_1} +\frac{\bar{W}^E\phi'(\phi^{-1}(p))}{\tau_1}+\lambda\right]\lambda(\lambda+1) +  \lambda\left[  \frac{\bar{W}^I \phi'(\phi^{-1}(p))\phi(\theta p) \theta}{\tau_1}\right] +(\lambda+1) \bar{I}^2 \frac{\phi'(\phi^{-1}(p))}{\tau_1\tau_2}\right)\bm I_{N_E}\nonumber \\& + & \lambda(\lambda+1)\frac{\bar{W}^E\phi'(\phi^{-1}(p))}{\tau_1} \bm 1_{N_E} \bigg) \label{step1}\\
&=& \det\left(-\hat{Q}(\lambda)\bm I_{N_E} +\lambda(\lambda+1)\frac{\bar{W}^E\phi'(\phi^{-1}(p))}{\tau_1} \bm 1_{N_E} \right)\nonumber
\end{eqnarray} 
where we have arrived at (\ref{step1}) by applying the matrix determinant identity:
\begin{eqnarray}
\det \begin{pmatrix}\bm A & \bm B \\  \bm C & \bm D \end{pmatrix} = \det\left(\bm A- \bm B\bm D^{-1}\bm C\right)\det \bm D
\end{eqnarray}
To proceed, we will note the following:
\begin{eqnarray}
\bar{W}^E\bm 1_{N_E} = \bar{W}^E\bm u \bm u^T, \quad \bm u  = \begin{pmatrix} 1\\1\\ \vdots \\1\end{pmatrix}
\end{eqnarray}
which allows us to use the rank-1 update to the determinant:
\begin{eqnarray}
\det\left(\bm A+\bm u\bm v^T\right)=\left(1+\bm u^T\bm A^{-1} \bm u \right)\bm \det(A\bm )
\end{eqnarray}
yields the following:
\begin{eqnarray}
C(\lambda) &=& (-1)^{N_E}\hat{Q}(\lambda)^{N_E} \left(1-\frac{\bar{W}^E\phi'(\phi^{-1}(p))}{\tau_1}\lambda(\lambda+1) \hat{Q}(\lambda)^{-1}\bm u^T \bm u \right)\\
&=& (-1)^{N_E}\hat{Q}(\lambda)^{N_E-1} \left(\hat{Q}(\lambda) -N_E\left( \frac{{W}^E}{N_E-1}\right)\frac{\phi'(\phi^{-1}(p))\lambda(\lambda+1)}{\tau_1}  \right)\\
&=& (-1)^{N_E} \hat{Q}(\lambda)^{N_E - 1}C_{SN}(\lambda)
\end{eqnarray}
where $C_{SN}(\lambda)$ is the characteristic polynomial for the single, recurrently coupled node.  Our analysis of the dual-node case applies here and the stability of the $N_E$-node system under mean-field assumptions applies.  The $N_E$ node system under mean-field connectivity has identical solutions to the single node system.  
\subsection*{The Normalized Excitatory Weight Solution} 
Finally, we will consider an arbitrary weight matrix, $\bm{W}^E$ with the only constraint being 
that 
\begin{eqnarray}
\sum_{j=1}^{N_E}\bm L^{EE}_{ij} = 1\label{ass1}, \quad \bm W^{EE} = W^E\bm L^{EE}.
\end{eqnarray}
  Note that the mean-field example previously considered is a special case of assumption (\ref{ass1}).   The assumption is required as it stabilizes the equilibria for $\bm W^I_{i}$  to $\bm W^I_i  = \bar{W}^I$ for all $i=1,2,\ldots N_E$ where $\bar{W}^I$ is the single-node equilibrium solution.   Using a similar derivation procedure as before, the characteristic polynomial simplifies to: 
\begin{eqnarray}
C(\lambda) &=& \det\left(-\hat{Q}(\lambda)\bm I_{N_E} +\lambda(\lambda+1)\frac{\phi'(\phi^{-1}(p))}{\tau_1} \bm W^{EE} \right)\label{dt1}\\
\hat{Q}(\lambda) &=& \lambda^3 + \lambda^2\left(\frac{1}{\tau_1}+1 \right) +  \lambda\left(\frac{1}{\tau_1} +\frac{ \overline{W^I} \phi'(\phi^{-1}(p))\phi'(\theta p)\theta}{\tau_1}+\frac{\overline{I}^2 \phi'(\phi^{-1}(p))}{\tau_1\tau_2}\right)\nonumber+\frac{\overline{I}^2 \phi'(\phi^{-1}(p))}{\tau_1\tau_2} \label{QL2}.
\end{eqnarray} 
This implies that
\begin{eqnarray}
C(\lambda) &=& \det\left(\lambda(\lambda+1)\frac{\phi'(\phi^{-1}(p))}{\tau_1} \bm W^{EE}-\hat{Q}(\lambda)\bm I_{N_E}  \right)\\
&=& \det\left(q_0(\lambda) \bm L^{EE}-\hat{Q}(\lambda)\bm I_{N_E}  \right), \quad q(\lambda) = (\lambda(\lambda+1)\frac{\phi'(\phi^{-1}(p))}{\tau_1}W^E \\
&=& q_0(\lambda)^{N_E} \det\left(\bm L^{EE} - \frac{\tilde{Q}(\lambda)}{q_0(\lambda)}\bm I_{N_E}\right)\\
&=& q_0(\lambda)^{N_E} \det\left(\bm L^{EE} -\mu I_{N_E} \right), \quad \mu= \frac{\tilde{Q}(\lambda)}{q(\lambda)}\\
&=& q_0(\lambda)^{N_E}\prod_{i=1}^{N_E} \left(\mu - r_i\right)
\end{eqnarray} 
where $r_i$ are the eigenvalues of $\bm L^{EE}$.  Undoing the substitutions resolves the factorized characteristic polynomial:
\begin{eqnarray}
C(\lambda) &=& \prod_{i=1}^{N_E} \left(\tilde{Q}(\lambda) -r_i  \frac{ \lambda(\lambda+1)\phi'(\phi^{-1}(p))}{\tau_1}W^E  \right)
\end{eqnarray} 
This factorization in $C(\lambda)$ allows one to resolve the Hopf bifurcation curve almost as easily as in the single node case.  In particular, if all the eigenvalues are real (for example, if the matrix $\bm W^{EE}$ is symmetric) then the Hopf bifurcation curve occurs when a complex conjugate pair of roots of
\begin{eqnarray}
\tilde{Q}(\lambda) -  \frac{r_i\lambda(\lambda+1)\phi'(\phi^{-1}(p))W^E}{\tau_1}
\end{eqnarray} 
cross $\text{Re}(\lambda_i) = 0$.  For the experimentally derived weight matrix we consider, all eigenvalues of $\bm W^{EE}$ are real due to the near symmetric nature of the matrix (\cite{Honey,Peter,hagmann}).   Thus, the potential Hopf-bifurcation curves are:
\begin{eqnarray*}
W^E_{Hopf,i}(\theta) &=& \frac{1}{r_i\phi'(\phi^{-1}(p))}\left(1-\tau_1 \mu_+(\theta)\right) \label{Hopf2}\\
\mu_\pm &=& \frac{-(D(\theta)+F(\theta)\kappa(\theta)+1-\kappa(\theta)) \pm \sqrt{ (F(\theta)\kappa(\theta)+D(\theta)+1-\kappa(\theta))^2 - 4 \kappa(\theta)F(\theta)(1-\kappa(\theta)}}{2(1-\kappa(\theta))}  \\
F(\theta) &=& \frac{1-r_ip^{-1}\phi^{-1}(p)\phi'(\phi^{-1}(p))}{\tau_1}\\
\kappa(\theta) &=& \frac{p\phi'(\theta p) \theta }{\phi(\theta p)r_i }\\
D(\theta) &=& \frac{\overline{I}^2 \phi'(\phi^{-1}(p))}{\tau_1\tau_2}.
\end{eqnarray*}
This implies that as we increase $W^E$, the first transition through $W^{E}_{Hopf,i}(\theta)$ yields a Hopf bifurcation.  Due to the form of (\ref{Hopf2}), this is likely to correspond to the eigenvalue of $\bm W^{EE}$, $\bm r^{max}_i$ with the largest positive real part.  Finally, we remark that if the row sum of the weight matrix $\bm L^{EE}$ is not constant but narrowly distributed around a mean-value $\bar{L}$, than one can readily derive the following approximation: 
\begin{eqnarray}
W^E_{Hopf,i}(\theta) &\approx& \frac{1}{r_i\phi'(\phi^{-1}(p))}\left(1-\tau_1 \mu_+(\theta)\right) \label{Hopf3}\\
r_i &=& \max_{i=1\ldots N_E}\{r_i\}\\
F(\theta) &=& \left(1-\frac{r_i}{\bar{L}}p^{-1}\phi^{-1}(p)\phi'(\phi^{-1}(p))\right)\\
\kappa(\theta) &=& \frac{p\phi'(\theta p) \theta \bar{L} }{\phi(\theta p)r_i }
\end{eqnarray} 
which we apply to the coupling matrix from \cite{Peter} where $\bar{L} = 0.2318$ and $r_i = 0.3148$.  
 
\end{document}